\input{psfig.sty}

\documentclass[useAMS,usegraphicx]{mn2e}
\usepackage{amsmath}
\def\xte{XTE J1751--305} 
\def\ltsima{$\; \buildrel < \over \sim \;$}
\def\simlt{\lower.5ex\hbox{\ltsima}}
\def\gtsima{$\; \buildrel > \over \sim \;$}
\def\simgt{\lower.5ex\hbox{\gtsima}}
\def\1751{XTE J1751--305 }  

\title[Measuring the spin up of the AMSP {\xte}]{Measuring the spin up of the Accreting Millisecond Pulsar {\xte}}

\author[A.Papitto, M.T.Menna, L.Burderi, T.Di Salvo and A.Riggio]{A. Papitto$^{1,2}$\thanks{E-mail: papitto@oa-roma.inaf.it}, M.T.Menna$^{2}$,  L.Burderi$^{3}$, T.Di Salvo$^{4}$, and A.Riggio$^{3}$\\ 
$^{1}$Dipartimento di Fisica, Universit\'a degli Studi di Roma ''Tor Vergata'', via della Ricerca Scientifica 1, 00133 Roma, Italy\\ 
$^{2}$Osservatorio Astronomico di Roma, via Frascati 33, 
Monteporzio Catone, 00040, Italy\\ 
$^{3}$Dipertimento di Fisica, Universit\'a degli Studi di Cagliari, SP Monserrato-Sestu, KM 0.7, Monserrato, 09042 Italy \\ 
$^{4}$Dipartimento di Scienze Fisiche ed Astronomiche, 
Universit\`a di Palermo, via Archirafi 36, Palermo, 90123, Italy}

\begin{document}

\maketitle 

\label{firstpage}

\begin{abstract}  We perform a timing analysis on RXTE data of the
accreting millisecond pulsar \1751 observed during the April 2002
outburst.  
After having corrected for Doppler effects on the pulse phases due 
to the orbital motion of the source, we performed a timing
analysis on the phase delays, which gives, for the first time for this
source, an estimate of the average spin frequency derivative 
$<\dot{\nu}>=(3.7\pm1.0) \times 10^{-13}$ Hz/s.  We discuss the torque resulting from the spin-up of the
neutron star deriving a dynamical estimate of the mass accretion rate and
comparing it with the one obtained from X-ray flux. Constraints
on the distance to the source are discussed, leading to a lower limit
of $\sim 6.7$ kpc.
\end{abstract}


\section{Introduction}

Accreting millisecond pulsars (AMSP in the following) 
are the long sought connection between low mass
X-ray binaries (LMXBs) and millisecond radio pulsars. In fact, although it was
hypothesised soon after their discovery that fast spinning radio pulsars were
``recycled'' by an accretion phase in a LMXB system, during which the neutron
star (NS) is spun--up (see for a review Bhattacharya \& van den Heuvel 1991), evidence has been
elusive since SAX J1808.4-3658, the first accretion-driven millisecond X-ray
pulsar, was discovered (Wijnands \& van der Klis 1998). SAX
J1808.4-3658, with a spin period of $2.5$ ms, exhibiting both X-ray bursts and
coherent pulsations, proved to be the missing link between the two classes of
sources.  Since then, six more millisecond X-ray pulsars were discovered (see Wijnands 2006 for an observational review).

All of these sources are transients with usually low duty
cycles. Except for the case of HETE J1900.1-2455 which remained active
for more than a year after its discovery in June 2005  (Galloway et
al. 2007), the outbursts of AMSP last for no more than a couple of
months, with recurrence times usually larger than $2$ yr (Galloway
2006). Although the sample is still small, monitoring of future
outbursts exhibited by the known sources is extremely important
for our understanding of LMXBs and their evolution.

The study of the rotational behaviour of these sources during outbursts is on
the other hand obviously fundamental as a test of theories of accretion
physics. As a matter of fact, the measure of the variations of the
spin frequency gives immediate understanding of the torques experienced by the
compact object because of the accretion of matter, and further allows a model
dependent dynamical estimate of the instantaneous mass accretion rate. Timing
techniques performed on the coherently pulsed emission (see e.g. Blandford \&
Teukolsky 1976) represent the key tool in order to directly measure the
variations of the spin rate of this kind of accretors. The application of this
kind of analyses on the X-ray emission of these sources, as observed by the
high temporal resolution satellite Rossi X-Ray Timing Explorer ({\it RXTE})
(Bradt et al. 1983), allowed the measurement of the spin frequency derivative
in the case of IGR J00291+5934 (Falanga et al. 2005; Burderi et al. 2007), XTE
J0929-314 (Galloway et al. 2002), XTE
J1814-338 (Papitto et al. 2007), XTE J1807--294 (Riggio et al. 2007) 
and of one of the oubursts shown by SAX
J1808.4--3658 (Burderi et al. 2006). See Di Salvo et al. (2007) 
and references therein for a review.

Even though the shortness of the outbursts generally exhibited by AMSP
strongly limits the capability of the timing analysis in discriminating
between various accretion models, observations have already shown
how the behaviour of these sources can be variable, in fact, as a result of
accretion, some of them are observed to  spin up, while others
decelerate. In this work we present a timing analysis performed on
the only outburst of {\xte} observed so far by {\it RXTE} with its highest
temporal resolution mode, and discuss the observed spin
frequency evolution as a result of the accretion torques acting on the
NS. Moreover the measure of the spin frequency derivative is used to infer a
dynamical estimate of the peak mass accretion rate during the
outburst, which can be compared with the estimate deduced from
spectral modelling of the X-ray emission, in order to give an estimate
of the distance to the source.

\section{Observations and Data Analysis}

{\xte} was first detected monitoring the
Galactic bulge region with the Proportional Counter Array (PCA) on
board the {\it RXTE} (Markwardt \& Swank 2002a). Subsequently, pointed observations in April 2002, 
allowed
the detection of an X-ray periodic modulation at a frequency of about
$435$ Hz, establishing that this source belongs to the class of
accreting millisecond X-ray pulsars (Markwardt et al. 2002b, M02
hereafter).  Throughout this paper we used a sample of {\it RXTE}
public domain data taken between 4 and 30 April 2002. The data we use
to perform the timing analysis are those collected by the Proportional
Counter Array (PCA, Jahoda et al. 1996), which is made by five units
(PCUs) sensitive in the $2-60$ keV band, for an overall collecting
area of $\sim 6250$ cm$^2$.  Except the observations of April 4, which 
were taken in Good Xenon configuration with a $1 \mu s$ time resolution 
($2$ s read out time and $256$ energy channels), 
all the other PCA data we considered are Generic Event   
(E\_125us\_64M\_0\_1s) with a temporal resolution of $122 \mu s$ and
$64$ energy channels. All the data were processed and analysed using the 
 HEASARC FTOOLS v.6.0.

 The source was first spotted by the PCA in the desired collecting mode
at $T_0=52368.653$ MJD, which represents the start time of 
observations throughout this paper. Thereafter the X-ray light curve of {\xte}
is made of two exponential decays with different e-folding
factors. The first one describes the first $8.5$ d of observation and can
be modelled with the function
$L_X(t)=L_X(\bar{T})exp[-(t-\bar{T})/\tau_d^{(1)}]$, where 
$L_X(\bar{T})$ is the
 luminosity at $\bar{T}=52369.644$ MJD and $\tau_d^{(1)}=7.2$ d (see Gierlinski \& Poutanen (2005),
GP hereinafter, for the $2-20$ keV light curve; see also
M02). Subsequently the light curve experiences a sharp break and can be 
described with a similar decay function this time with
$\tau_d^{(2)}=0.63$ d. The source then switches back to the
quiescence emission levels, $\sim 10$ d after the first available
observation. Considering a physically motivated spectral model (see the
discussion for further details) GP estimated for the  bolometric 
X-ray/$\gamma$-ray luminosity
attained by {\xte} during the outburst  $L_X(\bar{T})=2.7\times 10^{37}
d_{8.5}^2$ erg s$^{-1}$, where $d_{8.5}$ is the distance to the source
in units of $8.5$ kpc.

The technique we used in order to perform the timing analysis on the 
pulsed emission is extensively described in Burderi et
al. (2007, see also Papitto et al. 2007). We first corrected X-ray
photons arrival times to the Solar system barycentre considering the
best position available for this source from Chandra (M02). 
We then focused on the orbital modulation of the phases to derive
an accurate orbital solution. The evolution of the phases,
measured by folding $90$ s long intervals around the M02 estimate
 of the spin period, was modelled with Eq.1 of Papitto et
al. (2007), without considering the term owing to the position
uncertainties (see below). The orbital parameters we obtain, namely 
the projected semi-major axis measured in light ms, $a\sin i/c$, 
the orbital period, $P_{orb}$, 
the time of passage of the NS at the ascending node of the orbit
, $T^*$, and the 
eccentricity, $e$, are
listed in Tab.\ref{tab1}, which contains for comparison purposes 
also the ones reported by M02.  The two orbital solutions are in 
good agreement whitin the quoted uncertainties.

\begin{table}
\centering
\begin{minipage}{84mm}
\caption{Orbital and timing parameters of {\xte} }
 \label{tab1} 
\begin{tabular}{@{}lll}
\hline
 & M02 & This work \footnote{Numbers in parentheses, referring to our values,
 are the $90\%$ confidence level uncertainties in the last significant figure, while the
 ones referring to M02 are given at $3\sigma$ confidence level on the 
last significant digits. The same confidence levels are considered in giving upper limits on the eccentricity, $e$.} \\ 
\hline 
{\bf Orbital solution} & & \\
$a\sin i/c$ (lt-ms)          & $10.1134(83)$       &  $10.125(5)$           \\ 
$P_{orb}$ (s)                & $2545.3414(38)$     & $2545.342(2)$          \\
$T^*$ (MJD)                  & $52368.0128983(87)$ 
\footnote{ The value reported here is revised with respect to the one originally quoted in 
M02 (Markwardt et al. 2007). } 
&  $52368.0129023(4)$     \\ 
Eccentricity $e$             & $<1.7\times10^{-3}$ & $<1.3\times 10^{-3}\;$ \\ 
\hline
{\bf Timing Solution} & & \\
$\nu_0$ (Hz) & $435.317993681(12)$ & $435.31799357(4)$ \\
$<\dot\nu>$ (Hz/s) & $<3\times10^{-13}$ \footnote{ This upper limit is to 
be considered on the absolute value of $<\dot\nu>$.} & $(3.7\pm1.0)\times10^{-13}$. \\
$\dot\nu(\bar{T})$  (Hz/s) $\alpha=2/7$ & & $(5.6\pm1.2)\times10^{-13}$\\
\hline
\end{tabular}
\end{minipage}
\end{table}

The times of arrival of X-ray photons were then reported to the line of nodes 
of the binary system orbit considering our orbital solution. As the time over 
which any uncertainty on the orbital parameters may possibly affect the phases of 
the X-ray pulsations ($P_{orb}$) is much smaller than the time required for the 
spin frequency derivative to produce a significant effect, the two effects on 
residuals can be decoupled. The effect on pulse phases due to the remaining 
uncertainty on the orbital parameters, $\sigma_{\phi\:orb}$, 
can therefore be treated as a 
normally distributed source of error (see Eq.$3$ of Burderi et al. 2007 for 
an expression of $\sigma_{\phi\:orb}$). The final uncertainty, $\sigma_{\phi}$, on the phase residuals 
we consider in the following is then the squared sum between $\sigma_{\phi\:orb}$ 
and the statistical error arising from sinusoidal fitting of the pulse profiles.

By folding each light curve corrected for the orbital motion of the
system around our guess for the spin frequency $\nu_F$ (which initially 
was the M02 estimate), we could
detect coherent pulsations until MJD $52377.6$, $9$ d
after the first observation available, when the X-ray flux had 
become approximately one tenth of the peak flux.

\begin{figure}
\includegraphics[width=84mm]{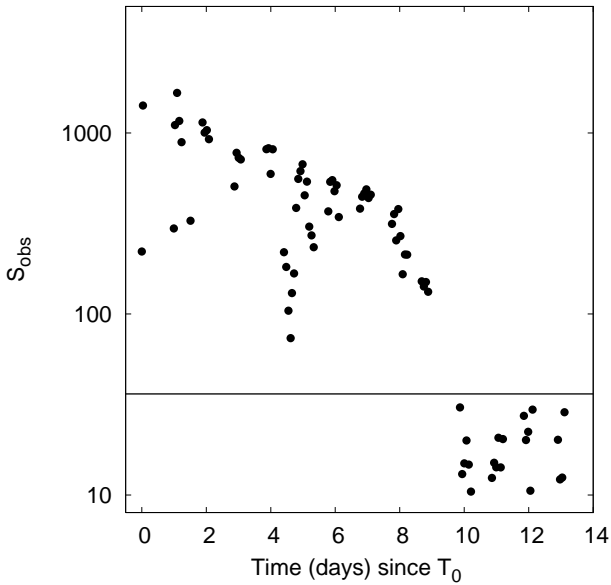}
\caption{ Plot of the observed statistics $S_{obs}$ of every considered folded light curve. The horizontal line represents the $99\%$ confidence limit for the detection of pulsations.}
\label{statistics}
\end{figure}

The criterion we considered in order to assess the presence of a
periodic signal is simply based on the
expected statistic distribution followed by a  folded light curve in which no
signal is present (Leahy et al. 1983). The statistics $S$ used
to check the presence of a periodic signal is defined as
\begin{equation}
\label{eq:statistic}  S=\sum_{j=1}^{n} \frac{(R_j-R)^2}{\sigma^2},
\end{equation}   where n is the number of phase bin by which the light
curve is divided, $j=1,...,n$, $R_j$ is the count rate in the j-th
bin, $R$ is the average count rate in the considered observation, and
$\sigma^2=R\times n/T$, with T total time of integration.  It is assumed 
that in the
absence of periodicity the photon counting in each bin of a folded light
curve, $R_j \times T/n$, is pure Poissonian counting noise, so that 
its mean and 
variance can be
estimated from the average number of counts in each bin, $R\times T/n$.
In the limit of a large
number of counts, as in the case of the 2002 outburst of {\xte}, when 
$R\times T/n \simgt 2000$, 
we can assume that this distribution behaves as a Gaussian with
the mean equal to the variance. If  these conditions are met the statistic
$S$ of a folded light curve with no signal is expected to follow a
chi-squared distribution with $n-1$ degrees of freedom.  It is then
immediate to define a threshold for the detection of the signal $S_0$
from the desired confidence level P,
$(1-P/100)=Q_{n-1}(\chi_0^2=S_0)$, where the term on the right hand
side is the integrated probability from $\chi_0^2$ to $\infty$ of a
chi-squared distribution with $n-1$ degrees of freedom. In this case
we chose $P=99$ and $n=19$, so that every observation with
$S_{obs}<36.19$ was considered as containing no signal at the
considered level of significance, and therefore withdrawn from the
sample used in timing analysis (see Fig.\ref{statistics}). 

\begin{figure}
\includegraphics[width=84mm]{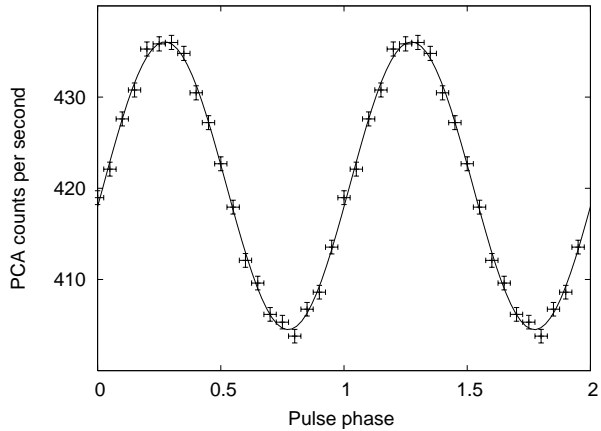}
\caption{Pulse profile produced folding in $20$ phase bins around  $\nu_0=435.31799357$ Hz, all the available observations spanning the time interval from MJD $52370.537$ to MJD $52370.776$ (OBSID 70131-01-02-00). This time interval does not represents a continuous time series and the effective total integration time is $\simeq 14$ ks. The solid line is the best fitting sinusoid. The reduced chi squared of this model is $\chi_r^2=0.78$ and is not significantly improved by adding an harmonic. Two cycles are plotted for clarity. }
\label{pulse}
\end{figure}

Folded light curves that met the detection criterion were then
modelled with a sinusoid with the period fixed to the folding one. We
also tried to add an harmonic to the fitting function, but such a
component was significantly detected and led to a slightly better
quality fit only in a small fraction of the observations ($\leq
10\%$), so that we consider for the purposes of the timing analysis
only the phases obtained with a single harmonic sinusoidal function.
An example of the pulse profile is shown in Fig.\ref{pulse}.

Therefore we modelled phase residuals  
with a parabolic function 
\begin{equation} \Delta\Phi(t)=1/\nu_F \times (\delta+\beta t+\gamma t^2),
\label{eq:residui}
\end{equation}  where $\nu_F$ is the folding frequency, 
$\beta=-\Delta\nu_0$ is the correction to $\nu_F$ to obtain an 
estimate of the spin frequency at $t=T_0$, $\gamma=-<\dot{\nu}>/2$ is 
the term owing to a constant spin frequency derivative and 
$\delta$ is a phase constant.
This procedure was repeated several times, correcting each time the frequency over 
which we fold the light curves, until the linear term of Eq.{\ref{eq:residui}} was 
compatible with being null within the uncertainties.
The timing parameters we finally obtain are $\nu_0=435.31799357(3)$ Hz and 
$<\dot{\nu}>=3.7(8)\times 10^{-13}$ Hz/s, where the numbers in parentheses are 
the $90\%$ confidence level uncertainties on the last significant figure, as 
for all the uncertainties quoted in the rest of the paper.

\begin{figure}
\includegraphics[width=84mm]{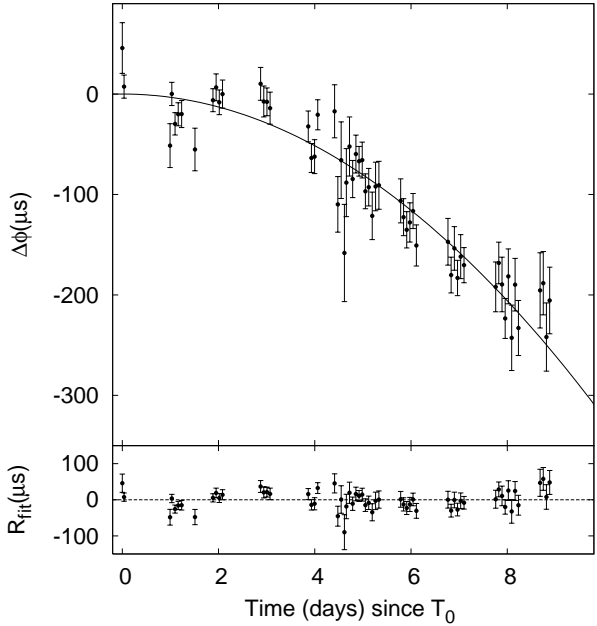}
\caption{Evolution of the pulse phase delays, measured in microseconds,
 folding
  every available observation around $\nu_0=435.31799357$ Hz.  The plotted 
  error bars are the $1\sigma$ uncertainties on phases, including both the 
  statistical errors from the pulse profile sinusoidal fitting and the errors
  induced by the residual uncertainties on orbital parameters. We note that these error bars have not been rescaled by the factor 1.2 that makes $\chi^2_{\rm r} = 1$, which we used to get the errors in the parameters given in Tab.{\ref{tab1}} (see text for details).
  The solid
  line is the best fit constant $\dot{\nu}$ model (upper
  panel). Residuals with respect to the best-fitting model
  (lower panel).} 
\label{timing}
\end{figure}

In Fig.\ref{timing} we plot the phase delays  $\Delta\Phi(t)$,  
measured in microseconds, of light curves folded
around $\nu_0$, versus the time elapsed since the first observation
considered.  The plotted error bars refer to the overall  $1\sigma$ 
uncertainties
on phases, computed by considering both the statistical errors coming
from the sinusoidal modelling and the errors induced by the
uncertainties on the orbital parameters listed in the right column of
Tab.\ref{tab1}, as already stated. 
The reduced chi squared of the quadratic best fitting model is $81/56$, which 
compared to the one obtained with a linear fit ($136/57$), gives an F test 
probability of $8\times10^{-8}$ that the improvement in the variance is purely given 
by chance. 

As already shown for other sources of this class (Burderi et al. 2006, Papitto et al. 2007, 
Riggio et al. 2007) 
the presence of timing noise, probably due to changes in the position of the emitting 
hot spot on the NS surface, suggests caution in considering the reliability of 
the uncertainties on the timing parameters obtained by a simple least square fit to 
the phase residuals, taken with their uncertainties $\sigma_{\phi}$. 
 Nevertheless the pulse phase evolution of {\xte} can be modelled satisfactorily by a 
general parabolic trend (see Discussion for a comparison with the behaviour displayed by 
other sources of this class), even if the presence of small timing residuals leads to a reduced 
$\chi^2$ slightly higher than one. Therefore, in order to get a conservative estimate of the uncertainties 
affecting the measured spin frequency and its derivative, we amplified all the errors 
on our phase measurements by a common factor ($1.2$) until we obtain $\chi^2_r=1$ from a fit 
with Eq.\ref{eq:residui}. We then consider as the most conservative 
the uncertainties $\sigma_{\nu_0}$ 
and $\sigma_{\dot{\nu}}$ so obtained. In this way we get to our final estimates of the timing parameters, 
which are the ones listed in Tab.{\ref{tab1}}. For sake of clarity we note however that the error bars plotted in Fig.{\ref{timing}} are not multiplied by any factor and 
they represent the genuine $1\sigma$ uncertainties on the measured phases.

A more physically motivated description of the evolution of phase delays is
easily obtained if the dependency of $\dot{\nu}(t)$ 
on the accretion rate is considered. 
According to the simplest model of angular
momentum exchange between the accreting matter and the compact object,
matter in the disc can be considered fully diamagnetic, so that the
only relevant torque is the positive one coming from accretion of
matter at the inner boundary of the disc, which is placed at a 
distance $R_{in}$ from the centre of the compact object.  
In an X-ray pulsar this
boundary can be determined, at least by orders of magnitude,
 considering the balance between the ram pressure of the
infalling matter and the magnetic pressure of the field that truncates 
the disc. When mass accretion rate declines, as in the fading
parts of the outburst light curves of X-ray transients like {\xte},
the pressure of falling matter is expected to decrease 
letting the magnetosphere to
expand.  Assuming that the inner disc boundary scales as $R_{in}
\propto \dot{M}^{-\alpha}$, with $\alpha\geq0$ and constant at every
 experienced accretion rate, the integral of the
spin up torque exerted by matter accreting at $R_{in}$ can be therefore 
carried on in terms of the instantaneous value of the mass accretion
rate. The instantaneous spin frequency derivative can then be expressed
as (see e.g. Burderi et al. 2007): 
\begin{equation}
\label{eq:torque}
\dot{\nu}(t)=  \frac{1}{2\pi I} l_0
\dot{M}(\bar{t}) \left(\frac{\dot{M}(t)}{\dot{M}(\bar{t})}\right)^{1-\alpha/2},
\end{equation}
 where $l_0=(G M
R_{in}(\bar{t}))^{1/2}$ is the specific angular momentum of accreting matter
evaluated at $t=\bar{t}$. Once it is assumed that $\dot{M}(t)\propto L_X(t)$, 
the temporal dependence of $\dot\nu(t)$ can be expressed in
terms of the particular shape of the outburst light curve. 
 As our aim is to compare our estimate of the accretion 
rate needed to explain the observed $\dot{\nu}$ assuming a certain torque 
model, and the accretion rate deduced from spectral fitting, we consider here 
and in the following $\bar{t}=\bar{T}$, an instant for which GP produce an 
accurate estimate of the $0.7-200$ keV flux.

Considering an X-ray light curve shape composed only by 
a single exponential decay, the
 expected behaviour of the phase delays time
evolution is then:
\begin{equation} 
\label{eq:phase_del}
\Delta\Phi(t)=\frac{1}{\nu_F}\left\{A-\left[B+k\tau_{\alpha}C\right](t-T_0)
-C\tau_{\alpha}^2\exp\left[-\frac{t-\bar{T}}{\tau_{\alpha}}\right]\right\},
\end{equation}  where $\tau_{\alpha}=\tau_d(1-\alpha/2)^{-1}$, 
$\tau_d$ is the e-folding factor of the light curve exponential decay, 
$k=\exp[-(T_0-\bar{T})/\tau_{\alpha}]$, A is a
phase constant, $B=\Delta\nu_0$ is the correction to the folded spin
frequency and $C=\dot\nu(\bar{T})=(2\pi I)^{-1}[G M R_{in}(\bar{T})]^{1/2}\dot{M}(\bar{T})$
is the spin frequency derivative at $t=\bar{T}$.

We first considered eq.(\ref{eq:phase_del}) with $\tau_d=7.2$ d, and
fitted the observed phase delays with various values of $\alpha$,
corresponding to different models of disc-magnetosphere
interaction. In particular, the case $\alpha=0$ corresponds to no
dependence of $R_{in}$ on the instantaneous accretion rate and in this
case we consider $R_{in}=R_C$ where $R_C$ is the corotation radius 
defined as the
radius at which the Keplerian orbiting matter in the disc exactly
corotates with the NS magnetosphere (see e.g. Rappaport et al. 2004).  On the
other hand, $\alpha=2/7$ is the Alfvenic value  obtained in the
approximation of spherical accretion of matter. Unfortunately we succeded in 
having no significant improvement in the description of the phase residuals by implementing 
such torque models, as for example we obtain $\chi^2_r=79/56$ 
when $\alpha=2/7$ is considered.

We also tried other $\alpha$ plausible values, based on a more realistic
treatment of the disc structure at the inner rim, which includes different
possible regimes, as gas pressure dominated or radiation pressure dominated SS
optically thick disc (respectively $\alpha=0.25$ and $0.15$ for models 1G and
1R in Ghosh 1996 and Psaltis \& Chakrabarty 1999).  We did not consider any
other model that would lead to significantly higher values of $\alpha$
(e.g. two temperature optically thin gas pressure dominated discs), as the
maximum value that this parameter can attain is the one that implies a full
excursion of $R_{in}$ during the outburst, from the NS radius to the
corotation radius, while the source keeps showing pulsations (see
discussion). Since in the case of {\xte} a coherent signal can be detected as
long as the X-ray flux, which is related to the instantaneous mass accretion
rate, has declined to one tenth of its maximum value, the constraint
$R_{in}(t_{cut})/R_{in}(T_0)=(F_x(t_{cut})/F_X(T_0))^{-\alpha} < R_C/R_{NS}$,
where $t_{cut}=T_0+9$ d is the time at which the pulsations are cut off, 
leads
to $\alpha<0.4$ for a $1.4\:M_{\odot}$ NS with a $R_{NS}=11.1$ km radius.  The
available statistics nevertheless revealed itself to be too low to
discriminate between these models. This is witnessed by the fact that the
goodness of the fit and also the value of the spin frequency
derivative remain almost the same when different plausible values of $\alpha$
are chosen. In order to give a reference value of $\dot{\nu}(\bar{T})$, the spin frequency derivative at $t=\bar{T}$, we consider
the simple Alfvenic case, $\alpha=2/7$, 
for which  
$\dot\nu(\bar{T})=(5.6\pm1.0)\times
10^{-13}$ Hz/s. In the same way we operated before, we amplified the actual errors on 
phases by a common factor until we have $\chi^2_{r}=1$ for the best fitting $\alpha=2/7$
 model, thus obtaining the $90\%$ confidence level uncertainty on  $\dot{\nu}(\bar{T})$ 
listed in Tab.{\ref{tab1}}.

We note that an attempt was also made to model the behaviour of the
phase delays in terms of the broken decay effectively observed in the 
X-ray light curve, rather
than considering a single decay taking place throughout the outburst.
We thus implemented in Eq.\ref{eq:phase_del} the change of the
e-folding factor that takes place simultaneously to the break. This
attempt anyway led neither to improvements in the quality of the fit
nor to variations in the measured parameters for every value of
$\alpha$ considered. This is probably due to the shortness of the time
interval in which the light curve is described by $\tau_d^{(2)}$
before the pulsations fade away.

We conclude this section by noting that a systematic term due to the 
uncertainty in the source position has to be
considered in order to have a reliable estimate of the  uncertainty
 in the spin frequency as well as in the spin frequency derivative. 
The best position available has a 90\% error circle
of $0".6$ (M02), and following the expression given by Burderi et
al. 2007, we estimate the upper limit on the effects of this uncertainty 
 on the value of $\nu_0$ as $\sigma_{syst\;\nu_0}<1\times10^{-7}$ Hz, while
on $\dot{\nu}(\bar{T})$ as $\sigma_{syst\;\dot{\nu}}<0.3\times10^{-13}$ Hz/s,
which is one order of magnitude smaller than the spin frequency derivative we measure,
 and smaller than the error on $<\dot{\nu}>$ quoted above.

\section{Discussion}

In the previous section we described the application of different
accretion models, referring to different assumption on the
disc-magnetosphere interaction and on the particular structure of the
inner region of the accretion disc, to the observed evolution of the
pulse phases of {\xte}, under the assumption that the  
curvature observed in the phase delays versus time 
is indeed a measure of the spin frequency derivative. 

The stability of the pulse phase evolution of this source resembles 
the one shown by 
IGR J00291+5934 (Burderi et al. 2007) and XTE J0929--314 (Galloway 
et al. 2007) with a smooth variation along the course of the outburst, witnessed 
by the small post fit residuals obtained 
even when a simple constant $\dot{\nu}$  model is used. This behaviour is 
different from the one analysed in an other subset of sources of this 
class, namely XTE J1814--338 (Papitto et al. 2007) and 
XTE J1807--294 (Riggio et al. 2007), where the phases oscillate around the mean 
trend clearly anticorrelating with rapid ($\sim 1$ d) X-ray flux 
 variations, and 
SAX J1808.4--3658 (Burderi et al. 2006), which shows an even more complex 
behaviour. It has to be noted that for all of the latter 
at least two harmonics are needed to model adequately the pulse profiles, 
while, among the sources of the first group, only XTE J0929--314 has a 
significant harmonic content.
 In particular the trend followed by the phases of the second harmonic seems 
more stable than the one of the fundamental (Burderi et al. 2006, Riggio et al. 2007), 
leading the authors to consider the second harmonic to establish the rotational behaviour 
of the considered source.
  A correlation 
between the phase stability and the two class of AMSPs depicted above 
seems to arise if it is considered that  for the 
last three the X-ray light curve is somehow complex, showing a variety of features, 
like re-flares, oscillations around a mean trend and in general 
variations on time scales of days, and simultaneously the pulse phases deviate from 
a continous trend. The light curves of the first three instead generally show   
a quite smooth exponential decay, during which the phases distribute themself normally 
around the mean trend.

A discussion of the noise of AMSP pulse phases around their mean trend is 
beyond of the scope of this paper, nevertheless 
these considerations make us even 
more confident 
 on the stability of the evolution of the pulse phases in the case 
of {\xte}, a source in which no secondary harmonics appear  nor the 
light curve deviates from an exponential decay through the time interval 
 considered here. We therefore ascribe the parabolic trend followed by 
the phases to a spin frequency evolution due to the accretion torques 
acting on the NS when it is efficiently accreting mass. 

However 
the attempts made proved unsuccessful in 
discriminating a constant spin up model from a model in which the 
spin frequency derivative depends on the instantaneous value of the 
mass accretion rate, as traced by the observed X-ray flux.
In the hypothesis that $\dot{\nu}(t) \propto \dot{M} (t) \propto F_X(t)$, which 
we stress is not favoured nor disfavoured by the data, the measurement of 
the spin frequency derivative of {\xte} at $t=\bar{T}$, 
$\dot{\nu}(\bar{T})$, allows a dynamical
estimate of the mass  accretion rate. This  can be further compared
with the estimate of the flux observed jointly by {\it XMM} and 
{\it RXTE}, in order to
constrain the distance to the source. The evaluation of
Eq.(\ref{eq:torque}) at the time $t=\bar{T}$, together with the estimate
of $\dot{\nu}(\bar{T})$ given in the previous section, leads to an expression
that relates mass accretion rate at that time to
the lever arm of the spin up torque at the same time, i.e. the inner
disc radius:
\begin{equation}
\label{eq:massaccr}
\dot{M}_0=(30\pm6)\times 10^{-10} I_{45} m^{-2/3} \xi^{-1/2}\:M_{\odot}/yr,
\end{equation} where $I_{45}$ is the moment of inertia of the compact
object in units of $10^{45}$ g cm$^2$, $m$ is the mass of the NS in solar
units, and $\xi=R_{in}/R_C$ is a parametrisation of the inner disc
radius at $\bar{T}$ in terms of the corotation radius $R_C$. For {\xte}
we have $R_C=(GM/\Omega_S^2)^{1/3}=26.1 m^{1/3} $ km, where
$M$ is the mass and $\Omega_S$ is the angular rotational velocity of
the neutron star. The use of such a parameter is particularly suited
for an accreting pulsar, as it has to be $R_{NS}/R_C<\xi(t)<1$ in
order for pulsations to be visible at a certain time $t$. The lower
constraint is an obvious consequence of the interpretation of coherent
pulsations as due to funnelling to the magnetic poles of the accreted matter,
while the upper limit is due to the centrifugal inhibition of
efficient accretion of matter onto the NS when the 
NS-magnetosphere system rotates faster than matter at the inner boundary of
the disc. Expressing these boundaries in terms of $m$ and $R_6$, the
radius $R$ of the NS in units of $10^6$ cm, one obtains for {\xte}, $0.38
R_6 m^{-1/3} < \xi < 1$, which allows the definition of a range of
possible accretion rates at $t=\bar{T}$, by their
substitution in Eq.(\ref{eq:massaccr}):
\begin{equation}
\label{eq:limit}
(30\pm6) I_{45} m^{-2/3} < \dot{M}_{10}(T_0) < (48\pm9) I_{45} R_6^{-1/2} m^{-1/2}
\end{equation}
where $\dot{M}_{10}$ is the mass accretion rate in units of $10^{-10}$ $M_{\odot}$/yr. 

This estimate of the mass accretion rate can be expressed in terms of
X-ray luminosity via the usual relation $L_X=\epsilon G M \dot{M} /
R_{NS}$, where $\epsilon\simeq1$ is the rate of conversion of
gravitational energy released in accretion to observable X-ray
luminosity. This allows to make a comparison of the dynamical estimate
of the accretion luminosity we derived from timing analysis with the
one obtained by spectral modelling of the observed X-ray flux, which
is obviously dependent on the source distance. By considering the
expression given by GP for the bolometric luminosity, $L_{X}(\bar{T})$,  
one obtains for the distance $d$ of {\xte}:
\begin{equation}
\label{eq:distance}
d=8.2\: I_{45}^{1/2} m^{1/6} R_6^{-1/2} \epsilon^{1/2} \xi^{-1/4} kpc
\end{equation}

Considering a moderately stiff EoS for an  $m=1.4$ NS, such as the FPS, one has for a compact object spinning at the rate measured for {\xte} 
$I_{45}=1.24$ and $R_6=1.11$ (see e.g. Cook, Shapiro \& Teukolsky
1994). It is then possible to find a range for the distance $d$ of {\xte}, 
$9.1 kpc \simlt d\:\epsilon^{-1/2} \simlt 11.6 kpc $.

{\xte} is located only $2^{\circ}$ away from the Galactic Centre, so that
a distance higher than $8.5$ kpc would be highly improbable,
if the source shows significant emission at energies of the order of
$\sim 1$ keV (GP estimated $N_H\sim 10^{22}$ $cm^{-2}$). However, the 
discrepancy between our lower limit on the distance, $9.1$ kpc, and the one 
implied by this constraint is not large, especially considering the
relatively large uncertainties underlying the simple arguments which led to 
our distance determination, and the assumptions made on the torque model 
and on the NS structure.  Moreover we recall that, because of the 
uncertainty in the source position, the estimate of 
$\dot{\nu}(\bar{T})$ we used  to get a dynamical estimate of the mass accretion rate is affected by a 
systematic error, 
which can in principle decrease our estimate of the distance below $8.5$ kpc.
 M02 estimated a lower limit on the distance of 
$7$ kpc based on indirect arguments, and our comparison between the observed $\dot{\nu}$ and the measured X-ray flux strongly supports the 
hypothesis that the source is close to the Galactic Centre.

 Nevertheless, we can argue that the limits on the distance 
we find above may be overestimated due to several reasons, among which, 
a non isotropic emission from the source or an occultation of part of 
the accretion luminosity.
In particular, as already noticed by GP, the fact that the pulse shape 
 shown by
{\xte} is almost sinusoidal implies that only one spot is visible
from our line of sight. If this had not  been the case, i.e. if
also an antipodal spot had intercepted our line of sight
during rotation, the folded light curves should have shown a secondary
maximum (or a plateau in a more extreme case), that would have required the
addition of at least an harmonic to be efficiently modelled. 
This picture is consistent with the results 
of detailed 3D magnetohydrodynamics simulations of disc accretion to 
this kind of rotators performed by Kulkarni \& Romanova (2005).
We argue that mass is accreting on both
the polar caps of the compact object, with a fraction of the emitted accretion luminosity 
being blocked by an opaque absorber (probably by the accretion disc, 
as the absence of a significant Compton reflection component 
strongly suggests $i>60^{\circ}$, see GP), and re-emitted outside the 
considered $0.7-200$
keV energy band (see also Burderi et al. 2007). It could be therefore the case that the value given by 
GP for 
the X-ray flux represents an
underestimate of the real emission owing to the accretion of matter.
This in turn would decrease our distance estimates to values in a better 
agreement with the constraint of the source not being farther than the 
Galactic Centre.
Defining $\eta$ as the ratio between the effective accretion luminosity 
and the one observed, we find lower distance estimates that would place 
the source not farther than the Galactic Centre for $\eta=1.2$, 
and a lower limit of $6.7$ kpc corresponding to $\eta=2$, the maximum
 value $\eta$ may reasonably attain.

It has to be highlighted that all these estimates 
of the source distance are derived by assuming that no negative torque
is acting on the NS, due e.g. to threading of the disc by the magnetic
field lines in regions where matter in the disc rotates slower than
the threading field lines, as it is assumed to explain the rotational
behaviour of other source of this class (see e.g. Papitto et
al. 2007). If this would not have been the case, higher estimates of
the mass accretion rate would be obtained, as the observed spin
frequency derivative would represent the balance between a negative
torque and the positive one due to accretion at $R_{in}$. This would
straightforwardly lead to higher estimates of the distance.

\section{Conclusions}

We performed a detailed timing analysis upon the coherent pulsations
shown by {\xte} during its 2002 outburst as observed by {\it
RXTE}. We could detect such a signal in the period 2002 April $4-13$,
i.e. in all observations during which the source emitted an X-ray flux
above the quiescent level.

After having corrected for the orbital effects on phase residuals, we then showed
how this source joins
SAX J1808.4--3658 (Burderi et al. 2006), IGR J00291+5934 (Falanga et al. 
2005; Burderi
et al. 2007) and XTE J1807--294 (Riggio et al. 2007) 
in the class of those AMSPs that exhibit a spin up in the
$10^{12}-10^{13}$ Hz/s range, with an average rate of $(3.7\pm1.0)\times10^{-13}$ Hz/s.

We applied different accretion torque models to the observed phase evolution, 
but we did not succeed in having a significant improvement of its
description with respect to a constant spin up model 
either when implying a dependence of the spin frequency
derivative on the instantaneous accretion rate, nor when 
discriminating between different scenarios of interaction
between the accretion disc and the rotating magnetosphere.
Applying an Alfvenic toque model we derived  
${\dot{\nu}}_0= (5.6\pm1.2)\times 10^{-13} $ Hz/s for the spin frequency 
derivative one day after the first observation available. 

The measured value of the spin frequency derivative implies a peak accretion
rate of at least $15\%$ of the Eddington limit.  Such a high accretion rate
indicates that {\xte}, as already noted for IGR J00291+5934 (Burderi et
al. 2007), accretes matter during outbursts at a much higher rate than that
usually considered typical (a few per cent of $L_{Edd}$) for AMSPs.

 The equality between the mass accretion rate
deduced from timing when a simple Alfvenic torque model is considered, 
and the one obtained from a spectral analysis of
the X-ray emission would place the source slightly farther than 
the Galactic Centre ($d\simgt 9.1$ kpc), although timing-based 
determinations of the distance may be affected by the large 
uncertainties on the assumed torque model.

We have also discussed the possibility that the mass accretion rate 
is a factor between $1$ and $2$ higher than the one inferred by the 
X-ray luminosity, if the emission of the antipodal polar
cap is not visible due to the occultation by a thick absorber and re-emission
of this energy is out of the considered band. An occultation of the secondary
cap emission is highly improbable to be due to the star itself, while
it seems more probable to owe to the accretion disc, as GP stated that
the absence of a significant Compton reflection in the emitted
spectrum strongly indicates $i>60^{\circ}$.
 Defining the parameter $\eta$ as the fraction of the
accretion luminosity effectively emitted by the NS 
 in the $0.7-200$ keV energy band, with respect to the one observed 
($\eta=1\div2$),
 we get a distance estimate $< 8.5$ kpc when $\eta\simgt1.2$, while a 
lower limit of $\simeq 6.7$ kpc is obtained in the case $\eta=2$.

We acknowledge the use of {\it RXTE} data from the {\it HEASARC} public 
archive. We also thank the anonymous reviewer for useful comments.
This work was supported by the Ministero della Istruzione 
della Universit\'a e della Ricerca (MIUR) national program 
PRIN2005 $2005024090\_004$.

\end{document}